\documentclass[useAMS,usenatbib]{mn2e}

\RequirePackage[dvips]{graphicx}

\usepackage{color}
\def\aap{\emph{A.\& A.}}

\def\apj{\emph{ApJ.}}

\def\mnras{\emph{MNRAS}}

\title[Minimum Variability Time Scales of Long and Short GRBs]{Minimum Variability Time Scales of Long and Short GRBs}

\author[MacLachlan et al.]{G. A. MacLachlan$^{1}$\thanks{E-mail:
maclach@gwu.edu (GAM)}, A. Shenoy$^{1}$, E. Sonbas$^{2,3}$, K. S. Dhuga$^{1}$, B. E. Cobb$^{1}$,
\newauthor T. N. Ukwatta$^{1,3,4}$, D. C. Morris$^{1,3,5}$, A. Eskandarian$^{1}$, L. C. Maximon$^{1}$,
\newauthor and W. C. Parke$^{1}$\\
$^{1}$Department of Physics, The George Washington University, Washington, D.C. 20052, USA.\\
$^{2}$University of Adiyaman, Department of Physics, 02040, Adiyaman, Turkey.\\
$^{3}$NASA Goddard Space Flight Center, Greenbelt, MD 20771, USA.\\
$^{4}$Department of Physics and Astronomy, Michigan State University, East Lansing, MI 48824, USA.\\
$^{5}$Department of Physics, University of Virgin Islands, Virgin Islands, USA.\\
}

\begin{document}

\maketitle

\label{firstpage}

\begin{abstract}
We have investigated the time variations in the light curves from a
sample of long and short Fermi/GBM Gamma ray bursts (GRBs) using an
impartial wavelet analysis. The results indicate that in the source
frame, that the variability time scales for long bursts differ from that
for short bursts, that variabilities on the order of a few
milliseconds are not uncommon, and that an intriguing relationship
exists between the minimum variability time and the burst duration.

\end{abstract}


\section{Introduction}\label{Introduction}

The prompt emission from Gamma-ray Bursts (GRBs) shows complex time profiles that 
have eluded a generally accepted explanation.
\citet{Fenimore2000} reported a correlation between variability of GRBs and the peak isotropic
luminosity. The existence of the variability-luminosity correlation suggests that the prompt emission light curves have
embedded temporal information related to the microphysics of GRBs. Several models have been proposed to explain
the observed temporal variability of GRB light curves. Leading models such as the internal shock model~\citep{Kobayashi97} 
and the photospheric
model~\citep{Ryde04} link the rapid variability directly to the activity of the central engine. Others invoke relativistic outflow mechanisms
to suggest that local turbulence amplified through Lorentz boosting leads to causally disconnected regions which in turn act as
independent centers for the observed prompt emission.  
Within more recent models,  both \citet{Morsony2010}  and \citet{Zhang11} argue that the temporal variability may show two different 
scales depending on the physical mechanisms generating the prompt emission.

In order to further our understanding of the prompt emission phase of GRBs and to explicitly test some of the key ingredients in
the various models, it is clearly important to extract the variability for both short and long gamma-ray bursts in a robust and unbiased manner. 
It is also clear that the chosen methodology should not only be mathematically rigorous but also be sufficiently flexible to apply to transient 
phenomena with multiple time scales and a wide dynamic range. A wide dynamic range is naturally provided by the bimodal separation of GRB duration occurring at $T_{90} = 2$ sec
as observed by~\citet{Kouveliotou93} to distinguish between long and short duration GRBs.  

In this paper, we extract variability time scales for GRBs using a method based on wavelets. 
The technique for such a temporal analysis is universal, and has the advantage over
Fourier analysis that transients and frequency correlations can be
more easily picked out in the data. Results presented herein were compared with \citet{Bhat12} who gave
pulse parameters for approximately 400 pulses obtained from 34 GRBs. It was shown by \citet{MacLachlan12}
that the minimum variablility time scale tracks the rise times of pulses very well for over three orders of magnitude in time scale.
The relation of minimum variability time scales to pulse parameters has been extended to four orders of magnitude by \citet{Sonbas12} who 
applied the present technique to analyze X-ray flares.

The time scales being investigated here have power densities very near to that of the noise in the data which makes extracting these time scales nontrivial. A somewhat older but still interesting discussion of extracting signal in a noisy environment can be found in ~\citet{Scargle82} and a more
recent discussion found in~\citet{Kostelich93}.
The technique we offer is not necessarily new but is different from previous published wavelet analyses~\citep{Fritz98,Walker00,Tamburini09,Anzolin10}\footnote{Note that \citet{Fritz98} analyzed optical data and  \citet{Tamburini09,Anzolin10} X-ray data from Cataclysmic Variables.}
in that we apply a number of modifications suggested by various authors~\citep{Addison02,Coifman95translation-invariantde-noising,Percival00,Strang97} 
as explained further in Sec~\ref{DataAnalysis}. 

The layout of the paper is as follows:
the source of the data is described in section 2;  the main aspects of the wavelet methodology are outlined in section 3;  in
section 4 we provide the details of the data analysis; in section 5 we present and discuss our main findings; finally, in
section 6, we summarize our conclusions.

\section[]{Data} \label{data}
The Gamma-Ray Burst Monitor (GBM) on board Fermi observes GRBs in the energy range 8\ keV to 40\ MeV. The GBM is composed of
12 thallium-activated sodium iodide (NaI) scintillation detectors (12.7 cm in diameter by 1.27 cm thick) that are
sensitive to energies in the range of 8 keV to 1 MeV, and two bismuth germanate (BGO) scintillation detectors (12.7 cm diameter
by 12.7 cm thick) with energy coverage between 200 keV and 40 MeV. The GBM detectors are arranged in such a way that they
provide a significant view of the sky \citep{Meegan09}.

In this work, we have extracted light curves for the GBM NaI detectors over the entire energy range (8 keV - 1 MeV, also
including the overflow beyond ~1 MeV). Typically, the brightest three NaI detectors were chosen for the extraction.
Lightcurves for both long and short GRBs were extracted at a time binning of 200 microseconds. The long GRBs were extracted
over a duration starting from 20 seconds before the trigger and up to about 50 seconds after the $T_{90}$ for
the burst without any background subtraction. For short GRBs, durations were chosen to be
20 seconds before the trigger and 10 seconds after the $T_{90}$. The $T_{90}$ durations were obtained from the Fermi GBM-Burst Catalog~\citep{Paciesas12}. 
%GCN circulars distributed by the GBM team and directly from the literature.

\begin{figure}
\includegraphics[width=84mm]{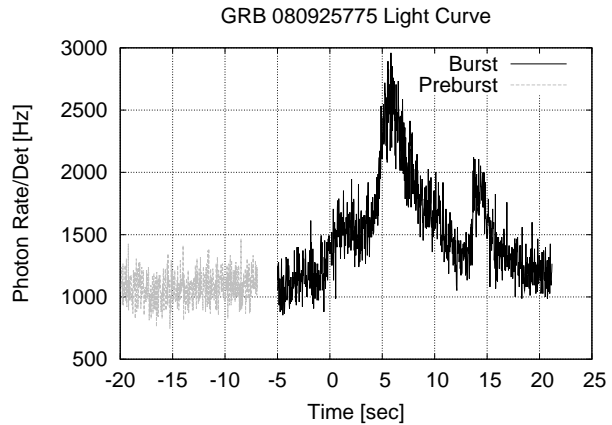}
\caption{GBM GRB080925775. Preburst portion of the light curve, used for background removal, is shown in gray. The burst portion, from
which a time scale is extracted, is shown in black.}\label{long_lc}
\end{figure}

\section[]{Methodology} \label{Methodology}

We report on our model-independent statistical investigation of the variability of Fermi/GBM long and short
GRBs. 
We extract this information by using
a fast wavelet transformation to encode GRB light curves into a wavelet representation and then compute a statistical measure of
the variance of wavelet coefficients over multiple time-scales.

\subsection{Minimum Variability Time Scales}

It is often the case when multiple processes are present 
that one process will dominate the others at certain
time scales but those same
processes may exchange dominance at other time scales.
A wavelet technique is useful in these situations because
the variances of wavelet coefficients 
are sensitive to
whichever processes
dominate the light curve at a given time
scale. Moreover, the technique
can be used to classify those dominant processes as well as
provide a means to determine the characteristic time scale,
$\tau_\beta$, for which
the processes exchange dominance.
Determination of $\tau_\beta$
helps in the
development of theoretical models and the understanding of observational data. 
Indeed, if there is a transition from a
time-scaling region to that of white noise 
%(down to the smallest time intervals in the data), 
then there is a smallest variability time for the physical processes present.

\subsection[]{Wavelet Transforms}\label{WaveletTransforms}
Wavelet transformations have been shown to be a natural tool for multi-resolution
analysis of non-stationary time-series \citep{Flandrin89,Flandrin92,Mallat89}.
Wavelet analysis is similar to
Fourier analysis in many respects but differs in that wavelet basis functions
are well-localized, \emph{i.e.} have compact support, while Fourier basis functions are global.
Compact support means that outside some finite range the amplitude of wavelet basis
functions goes to zero or is otherwise negligibly small \citep{Percival00}. 
In principle, a wavelet expansion forms a faithful representation of the original
data, in that the basis set is orthonormal and complete.

\subsubsection{Discrete Dyadic Wavelet Transforms}

Given the discrete nature of the data, we employ a discrete wavelet analysis.
The rescaled-translated nature of the wavelet basis functions make the wavelet transform
well-localized in both frequency and time, which results in an insensitivity
to background photon counts expressed by polynomial fits.
%polynomial backgrounds for photon counts. 
The level of insensitivity,
formally known as the vanishing moment condition, can be adjusted by
the choice of wavelet basis function.
By construction, the discrete wavelet transform is a multi-resolution operation \citep{Mallat89}.
Such wavelets, denoted $\psi_{j,k}(t)$, form a dyadic basis set,
i.e. wavelets in the set have variable widths and variable central time positions. 

The wavelet analysis employed in this study, as with the fast Fourier transform, 
begins with a light curve with $N$ elements,
\begin{equation}
X_i=\{X_0 \mathellipsis X_{N-1}\},
\end{equation}
where $N$ is an integer power of two. The light curve is convolved with a
scaling function,
$\phi_{j,k}(t_i)$, and wavelet function, $\psi_{j,k}(t_i)$ which are
rescaled and translated versions of the original scaling and wavelet functions
$\phi(t_i)=\phi_{0,0}$, and
$\psi(t_i)=\psi_{0,0}$.
Translation is indexed by $k$ and rescaling is indexed by $j$. The rescaling and translation relation is
given by
\begin{equation}
\psi_{j,k}(t) = 2^{-j/2}\psi(2^{-j}t-k).
\label{eq:dyad_wavelet}
\end{equation}

The precise forms of the scaling and wavelet functions are not unique.
The choices are made according to the features one wishes to exploit~\citep{Percival00,Addison02}.
The scaling function acts as a smoothing filter for the input time-series
and the wavelet function probes the time-series for detail information at
some time scale, $\Delta t$, which is twice that of the finest binning of the
data, $T_{\rm bin}$.
In the analysis, the time scale is doubled $$\Delta t \rightarrow 2\Delta t$$
and the transform is repeated until $$\Delta t = NT_{\rm bin}.$$

In this analysis we choose the Haar~\citep{Addison02} scaling/wavelet basis because it
has the smallest possible support, has one vanishing moment, 
and is equivalent to the Allan variance~\citep{Howe95}, allowing for a
straightforward interpretation.

\subsubsection{The Haar Wavelet Basis}
Convolving the light curve, $X$, with the scaling functions yields approximation coefficients,
\begin{equation}
a_{j,k} = \langle \phi_{j,k}, X \rangle.
\end{equation}

Interrogating $X$ with the wavelet basis functions yields scale and position dependent detail coefficients,
\begin{equation}
d_{j,k} = \langle \psi_{j,k}, X \rangle,
\end{equation}
It is interesting to point out that for the trivial $N=2$ case the Haar wavelet transform and the Fourier transform are identical.

\subsection{Logscale Diagrams and Scaling}\label{sec:LD}

Logscale diagrams are useful for identifying scaling and noise regions. 
Construction of a logscale diagram for each GRB proceeds from the variance of detail coefficients \citep{Flandrin92},
\begin{equation}
\beta_j = \frac{1}{n_j}\sum_{k=0}^{n_j-1}|d_{j,k}|^2,
\label{eq:waveletVariances}
\end{equation}
where the $n_j$ are the number of detail coefficients at a particular scale, $j$.
A plot of $\log_2$ variances versus scale, $j$, takes the general form
\begin{equation}
\log_2 \ \beta_j  = \alpha j+{\rm constant},
\end{equation}
and is known as a logscale diagram. A linear regression is made of each logscale diagram and the slope parameter, $\alpha$,
(depicting a measure of scaling) is estimated.
White-noise processes appear in logscale diagrams as flat regions while non-stationary
processes appear as sloped regions with the following condition on the slope parameter, 
$\alpha>1$~\citep{Abry03,Percival00,Flandrin89}. 

%Glen: I added the last line.  wcp

\section[]{Data Analysis} \label{DataAnalysis}

\subsection{Background Subtraction}
\label{cleanup}
We now present a method for removing photometric background due to
noise not intrinsic to the GRB so that physical variability arising
from the GRB remains for further analysis.
Background subtraction for a statistical analysis of variability via wavelet transforms should proceed in the space variances as
opposed to a traditional flat or linear subtraction of counts. This owes to the fact that Haar detail
coefficients are insensitive to polynomial trends in the signal up to first order. Subtraction of a flat or linear background
from a light curve is an operation under which the wavelet transform is invariant (as are Fourier transforms) 
apart from the mean signal coefficient. 

The GRB light curves show power at various variablity time scales. Most often, there
is a region of the logscale diagram (log-power verses log-varibility time) with a single slope,  
indicating scaling in the power over those variability-times,
and a flat region at the shortest variability times, indicating the presence of white-noise.  Some of this
white-noise may be intrinsic to the GRB.  Some may be attributed to 
instrumental noise and to 
background emissions from sources not including the GRB in question. 
We therefore express the variability of the burst, $\beta_j^{\rm burst}$, at time scales $j$ as comprising of individual
variances: a scaling component, $\beta^{\rm scaling}$; an intrinsic noise component, $\beta^{\rm noise}$; and a background
component, $\beta^{\rm background}$. The variability of the
burst can then be described as a linear combination of the component variances so long as the components have vanishing covariances.
In this event we write,
\begin{equation}
\beta_j^{\rm burst} = \beta_j^{\rm scaling}+\beta_j^{\rm noise}+\beta_j^{\rm background}.
%+\beta_j^{\rm bootstrap}
\label{eq:sumOfVariances}
\end{equation}
The minimum variability time scale, $\tau_\beta$, is identified from a logscale diagram by the octave, $j_{\rm {intersection}}$, of the intersection of the
flat intrinsic noise domain, $\beta_j^{\rm noise}$, with the sloped scaling domain, $\beta_j^{\rm scaling}$,
\begin{equation}
\tau_\beta\equiv T_{\rm bin}\times 2^{j_{\rm intersection}}.
\label{eq:mts}
\end{equation} 
In practice, the octave at which the intersection occurs is determined by equating the polynomial fits to the flat intrinsic noise domain and
the sloped scaling domain and solving for $j_{\rm intersection}$. The uncertainty in $\tau_\beta$ is determined by propagating the uncertainty in the 
parameters from the fits to the $\beta_j$ which in turn follow from a bootstrap procedure described in Sec.~\ref{circperm} and Sec.~\ref{sec:rtc}.
It is at this time scale, $\tau_\beta$, that a structured
physical process appears to give way to one that is stochastic and unstructured. Clearly one seeks to remove $\beta_j^{\rm background}$ from
Eq.~\ref{eq:sumOfVariances} to arrive at the cleanest possible signal,
\begin{equation}
\beta_j^{\rm burst}\rightarrow\beta_j^{\rm clean}\equiv\beta_j^{\rm burst} - \beta_j^{\rm background} = \beta_j^{\rm scaling}+\beta_j^{\rm noise}.
\label{eq:cleanedVariances}
\end{equation}

In order to estimate the variance of the background during the burst, we will assume that the variance
obtained from a preburst portion of the light curve can serve as an acceptable surrogate for the background variance. That is,
\begin{equation}
\beta_j^{\rm preburst} \equiv \beta_j^{\rm background},
\label{eq:backgroundSurrogate}
\end{equation}
and then the background is removed from the signal according to the relation,
\begin{equation}
\log_2(\beta_j^{\rm clean} ) = \log_2(\beta_j^{\rm burst}-\beta_j^{\rm preburst} ).
\label{eq:correction1}
\end{equation}
A simple algebraic manipulation of Eq.~\ref{eq:correction1} gives a form,
\begin{equation}
\log_2(\beta_j^{\rm clean} ) = \log_2(\beta_j^{\rm burst} ) + \log_2\left(1-\frac{\beta_j^{\rm preburst} }{\beta_j^{\rm burst} }\right).
\label{eq:correction2}
\end{equation}
For long GRBs, the preburst is defined relative to a 0 s trigger time as T-20 s to T-5 s and for short GRBs the preburst is defined to be
from T-15 s to T-1 s. Here T is the trigger time of the burst.

\subsubsection{Statistical Uncertainties}
We have considered the statistical uncertainties in the light curve by a typical bootstrap approach in which the square root of the number of counts per bin is used
to generate an additive poisson noise. A new poisson noise is considered for each iteration through the bootstrap process. 
More significant contributions to the uncertaintaies are discussed in Sec.~\ref{circperm} and Sec.~\ref{sec:rtc}.

\subsubsection{Circular Permutation}
\label{circperm}
Spurious artifacts
due to incidental symmetries resulting from accidental misalignment \citep{Percival00,Coifman95translation-invariantde-noising}
of light curves with wavelet basis functions are minimized by circularly
shifting the light curve against the basis functions. Circular shifting
is a form of translation invariant de-noising~\citep{Coifman95translation-invariantde-noising}.
It is possible a shift will introduce additional artifacts
by moving a different symmetry into a susceptible location. Thus, our approach is to circulate the signal through all possible values,
or at least a representative sampling, and then take an average over the cases which do not show 
spurious correlations.

\subsubsection{Reverse-Tail Concatenation}
\label{sec:rtc}
Both discrete Fourier and discrete wavelet transformations
imply an overall periodicity equal to 
the full time-range of the input data.  
This can be interpreted to mean that for a series of $N$ elements,
$\{X_0,X_1\mathellipsis X_{N-1}\}$ then $X_0$ is made a surrogate for $X_N$ and $X_1$ is made a surrogate for $X_{N+1}$, and so forth.
This assumption may lead to trouble if $X_0$ is much different from $X_{N-1}$.  In this case, artificially large
variances may be computed. 
Reverse-tail concatenation minimizes this problem by making a copy of the series
which is then reversed and concatenated onto the end
of the original series resulting in a new series with a length twice that of the original.
Instead of matching boundary conditions like,
\begin{equation}
X_0, X_1,\ldots,X_{N-1}, X_0,
\end{equation}
we match boundaries as,
\begin{equation}
X_0, X_1,\ldots X_{N-1},X_{N-1}, \ldots , X_1, X_0.
\end{equation}
Note that the series length has thus artificially been increased to $2N$ by reversing and doubling of the original series.
Consequently, the wavelet variances at the largest
scale in a logscale diagram reflect this redundancy. This is the reason the wavelet variances at the
largest scale are excluded from least-squares fits of the scaling region.

Another difficulty in wavelet expansions is that the initialization procedure of the multi-resolution algorithm may 
pollute the detail coefficients at
the finest scale \citep[see][]{Strang97,Abry03}. For this reason we follow the advice of~\citet{Abry03} 
and discard the detail coefficients at the finest scale.

\subsection{Simulation}
\label{sim}
\begin{figure}
\includegraphics[width=84mm]{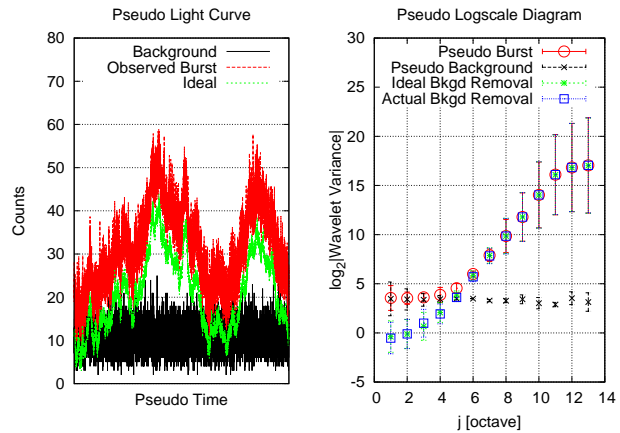}
\caption{In the left hand panel are simulated light curves and noise processes: an \emph{ideal} fBm process (green) and a 
\emph{background} Poisson noise (black). 
The sum of the fBm and Poisson processes is shown in red and is labeled \emph{observed}. The \emph{observed} light curve (red) is the
sum of the fBm and Poisson noise.
The right hand panel shows the 
results of the background subtraction procedure. Red and black points show the logscale diagrams corresponding to the \emph{observed} light curve 
and \emph{background}, respectively. The green data shows the logscale diagram for the \emph{ideal} light curve and the blue data
are the logscale with background removed. The agreement between green and blue data demonstrates the merit of the background 
removal procedure. }
\label{fig:simulation}
\end{figure}
\begin{figure}
\includegraphics[width=84mm]{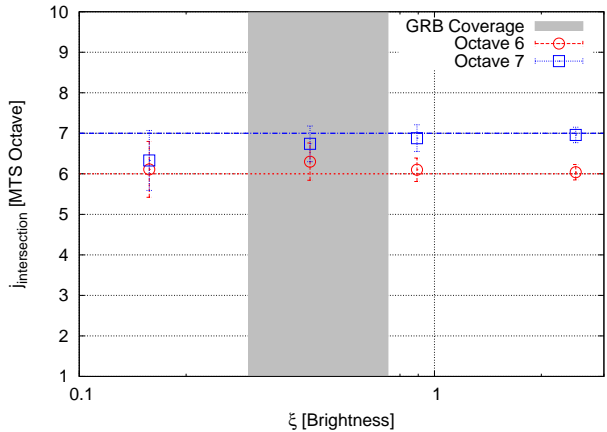}
\caption{Results of signal to noise sensitivity test.  We generated 1000 simulated light curves with expected $j_{\rm{intersection}}$ equal to 6 and 7 for various brightness. 
We show in gray the region along the $\xi$-axis 
where we find the GRBs analyzed in this paper, $0.3<\xi<0.74$. We note that
in the region covered by the GRBs presented herein, brightness does not affect
$j_{\rm{intersection}}$ greatly.
}
\label{fig:brightness_simulation}
\end{figure}

The efficacy of this background subtraction method and the sensitivity to 
signal to noise was tested using simulated data in the form of
fractional Brownian motion (fBm) time series that were first discussed by \citet{Mandelbrot68}.
One advantage of using fBms for simulation of time series data 
is that short duration, statistically significant fluctuations which 
trigger the identification of a 
minimum variability time scale arise naturally as part of the random process 
which produces them. Another is that fBms have a scaling parameter, 
$\alpha$, which is easily varied. 

The outline of the simulation procedure begins with using 
the numerical computing environment MATLAB
to produce 1000 realizations of fBms with scaling parameter $\alpha$ randomly chosen from the range $1.0\leq\alpha\leq2.0$
by using a uniform random number generator. 
The fBms were then acted upon by a Poisson operator which 
transformed each time series into a Poisson-distributed series but left other
properties of the fBm intact, e.g., $\alpha$. 

The fBms were then combined with a
Poisson noise with variance, $\lambda_B$.
These Poisson noises were regarded as intrinsic to the GRB.
Another set of Poisson noises with variances, $\lambda_I$, were generated 
and these noise signals were interpreted as external \emph{background}
meant to be removed by the subtraction procedure.

The idealized light curves were then combined with external backgrounds resulting in pseudo-\emph{observed} light curves
shown in red on the left panel of Fig.~\ref{fig:simulation}. The pseudo-observed light curves and
the external background noises were transformed into wavelet coefficients and wavelet variances were computed according to
Eq.~\ref{eq:waveletVariances}. The variances of the pseudo-observed light curve (labeled \emph{actual}) and the background are
plotted in the right panel of Fig.~\ref{fig:simulation} in red and black, respectively. The background was subtracted from the
pseudo-observed light curve as detailed in Eq.~\ref{eq:correction2} and the resulting corrected variances are plotted in blue in the
right panel of Fig.~\ref{fig:simulation}. The corrected variances are to be compared to the variances of the ideal light curve
which are plotted in green. 

\paragraph{Signal to Noise Sensitivity}
We define a brightness parameter, $\xi$, such that 
\begin{equation}
\xi\equiv 2^{-\delta/2},
\end{equation}
where
\begin{equation}
\delta\equiv\log_2 \lambda_B-\log_2 \lambda_I.
\end{equation}
Ideally the octave, $j_{\rm{intersection}}$, which is related to the minimum variability time scale, $\tau_\beta$, as defined in 
Eq.~\ref{eq:mts} is completely determined by 
$\xi$, $\lambda_I$, $\alpha$, and the 
standard deviation of the increments of the fBm, $\sigma$. 
We fixed the values of $\lambda$, $\sigma$, and varied 
$\alpha$ and $\xi$ so that
the expected values of $j_{\rm{intersection}}$  are $\langle j_{\rm{intersection}}\rangle=\{6,7\}.$ The time series
thus produced were then analyzed as described in Sec.~\ref{cleanup}. 
The results of the simulation are given in Fig.~\ref{fig:brightness_simulation}.
The horizontal red and blue lines show the expected $j_{\rm{intersection}}$ for 
octaves 6 and 7, respectively, and are given as a guide. Brightness, $\xi$, 
increases to the right. We show in gray the region along the $\xi$-axis 
where we find the GRBs analyzed in this paper based on the background noise 
level and the estimated noise level instrinsic to the GRB. We have noted 
by our own experience that for 
$\xi<0.1$ the technique discussed in this paper performs poorly. However, in 
the range of signal to noise sampled by the GRBs used here, $0.3<\xi<0.74$, the 
background subtraction technique does not suffer from a large 
systematic response to variations in 
brightness, as can be seen in Fig.~\ref{fig:brightness_simulation}.  

\paragraph{Flux Sensitivity}
We also investigated the reliability of the analysis as a function of flux by removing randomly selected counts from the original
simulated signal component while leaving the background noise level undisturbed.The analysis is then repeated for the newly \emph{dimmed} simulated light 
curves
and comparision is made to the original un-dimmed version. This brightness comparison is similar to the one described in~\citet{Norris95} but with a different normalization. In~\citet{Norris95} light curves were normalized by peak intensities. In this study the simulated light curves were normalized by signal power at the time scales specified by the dyadic partitioning of the wavelet transform.

Dimming of the simulated light curves was done by removing 0-10\% and reanalyzing then repeating by removing 10-20\% and so forth up to 70-80\%. We also considered the effect of larger variations in count removal, i.e., removing 0-25\% 25-50\% and 50-75\%.  
 We find that a decrease in flux has essentially the same effect on $\tau_\beta$ as increasing the noise level. However, the largest effect was by the wider bite of counts. For example, one can expect more accurate results from this analysis by removing 30-40\% of counts on a bin to bin basis than by removing 25-50\%. We conclude that flux related effects are more serious when it varies widely throughout the duration of the light curve. In all cases we have studied removing 80\% or more of the counts in the signal was
a reliable way to make the method fail. Fortunately when the method fails in this way it does so in an obvious way, i.e., the white noise signal power coefficients in the logscale diagram become highly irregular. 
As discussed in Sec.~\ref{select}, we test for this effect in the GRB data by performing a chi-squared test on the white noise signal power coefficients and rejecting any GRB that fails.      
 
In summary, 1000 simulated light curves were generated and background noise was added. The light curves with background noise
were then denoised using the same algorithm applied to actual GRB data in which preburst data were used as a surrogate for background.
The simulated background subtracted variances were then compared to the variances of the ideal light curves, i.e., light curves
without external background noise.
Signal to noise effects on the reliability of the method were also considered 
and found either to be small compared to our quoted errors or large enough that $\tau_\beta$ could not be determined. In the case of the latter 
the GRB was removed from the analysis. 
The results indicate that the background subtraction 
method is robust and gives confidence that external
background noise can be subtracted from the GRB light 
curves with the assumption that preburst data can serve as a 
surrogate for
background noise.

\begin{figure}
\includegraphics[width=84mm]{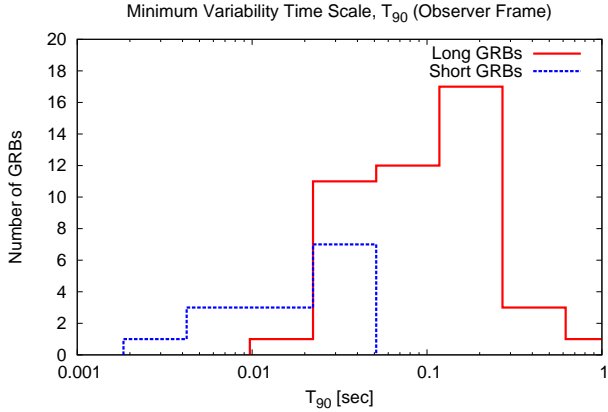}
\caption{A histogram of minimum variability time scales, in the observer frame, for long and short GRBs. 
It is clear that the distribution of long GRBs is displaced from the distribution of short GRBs.}
\label{fig:tau}
\end{figure}

\subsection{Selection Criteria}
\label{select}
\begin{figure}
\includegraphics[width=84mm]{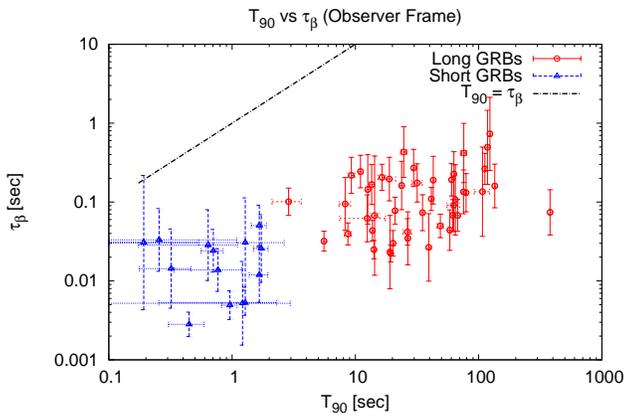}
\caption{Minimum variability time scale versus $T_{90}$ in the Observer frame.}
\label{TauVT90Obs}
\end{figure}

\begin{figure}
\includegraphics[width=84mm]{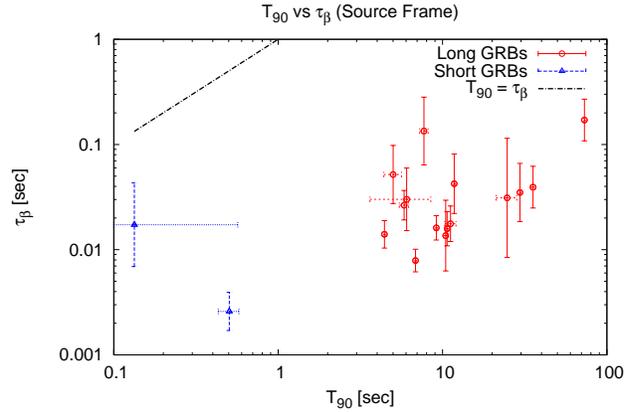}
\caption{Minimum variability time scale versus $T_{90}$ in Source frame. The correction for time dilation shortens $T_{90}$
and decreases the minimum variability time scale of each burst. }
\label{TauVT90}
\end{figure}

\begin{figure}
\includegraphics[width=84mm]{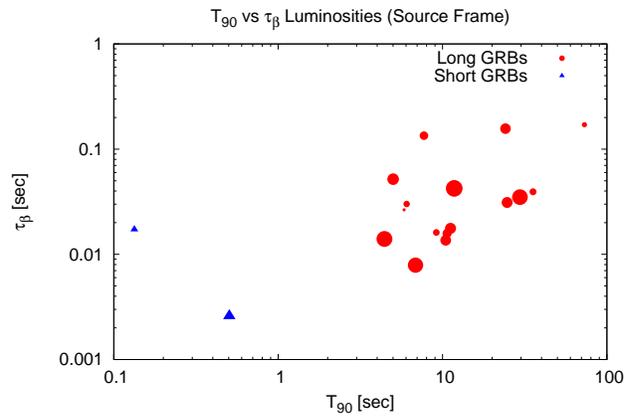}
\caption{Minimum variability time scale versus $T_{90}$ with symbol size determined by luminosity (larger symbols for higher luminosity). 
No obvious relation between minimum variability time scale and luminosity is apparent. See Fig.~\ref{TauVT90} for error bars.}
\label{TauVT90Lum}
\end{figure}

\begin{figure}
\includegraphics[width=84mm]{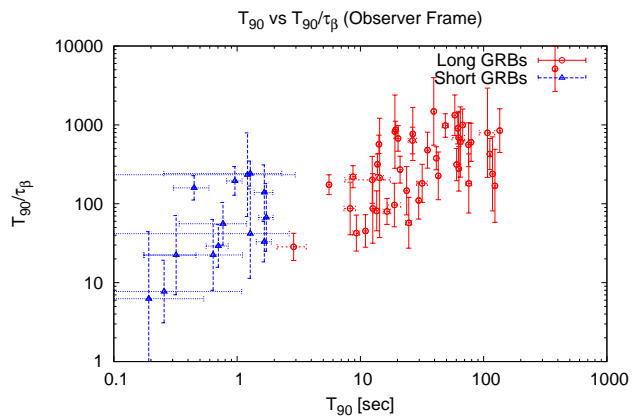}
\caption{The ratio of duration-to-minimum variability time scale ($T_{90}/\tau_\beta$) versus $T_{90}$.  }
\label{T90BYTauVT90Obs}
\end{figure}

We analyzed 122 GRBs (61 long and 61 short) listed in the Fermi GBM-Burst Catalog~\citep{Paciesas12}
for the first two years of the GBM mission. 
As discussed in Sec.~\ref{sim}, the signal-to-background ratio is a factor to be considered in recovering the intrinsic light curve
(see Eq.~\ref{eq:correction2}).
We required the following condition on the ratio of variances, 
\begin{equation}
\frac{ \beta_j^{\rm preburst} }{\beta_j^{\rm burst} } < 0.75,
\label{eq:secondpasscut}
\end{equation}
for one or more octaves, $j$.
In addition, we also required that the first order 
polynomial fits to the noise region and to the scaling region each had a $\chi^2$/d.f. that was less than 2. 
This reduced the sample to 14 short GRBs (Tab.~\ref{table:shorts}) and 46 
long GRBs (Tab.~\ref{table:longs}) for a total of 60 and it is 
these GRBs which are used to create Figs.~\ref{fig:tau},~\ref{TauVT90Obs}, and~\ref{T90BYTauVT90Obs}.
For boosting into the source frame (Figs.~\ref{TauVT90} and~\ref{TauVT90Lum}) 
a known $z$ is obviously required and this cut further reduced the data set to 
2 short GRBs and 16 long GRBs for a total of 18 GRBs considered in the source frame (see Tab.~\ref{table:good}). 

\section{RESULTS and DISCUSSION}

For a large sample of short and long GBM bursts, we have used a technique based on wavelets to determine the minimum time scale 
($\tau_\beta$) at which scaling processes dominate over random noise processes. 
The $\tau_\beta$ is the intersection of the scaling region (red-noise) 
of the spectrum in the logscale diagram with that of the ‘flat’ portion representing the (white-noise) random noise component. 
This transition
time scale is the shortest resolvable variability time for physical
processes intrinsic to the GRB.
Histograms of the extracted $\tau_\beta$ values for long and short GRBs are shown in Fig.~\ref{fig:tau}. 
We make two 
observations regarding these histograms: (1) There is a clear temporal offset in the extracted mean $\tau_\beta$  
values for long and short GRBs. 
We believe this is the first clear demonstration of this temporal difference. \citet{Walker00}, who studied the temporal 
variability of long and short bursts using the BATSE data set did not report a systematic difference between the two types of bursts. 
(2) The two histograms are quite broad and very similar in dispersion. While the difference in the mean $\tau_\beta$  is understandable 
(a point we discuss further elsewhere) the similarity of the dispersion is somewhat surprising since the progenitors and the 
environment for the two types of bursts are presumably very different. The comparison is qualitative at best however because 
the $\tau_\beta$  scale has not been corrected for redshift ($z$), an effect that impacts the long bursts more than the short bursts. 
In passing we note that the dispersion of the $\tau_\beta$  
histogram (for long bursts) is in agreement with the results of \citet{Ukwatta11} 
who performed a power density spectral analysis of a large sample of Swift long GRBs. In that work the authors extracted threshold 
frequencies and related them to a variability scale.

In Fig.~\ref{TauVT90Obs} 
we show a log-log plot of $\tau_\beta$  versus $T_{90}$ (the duration of the bursts); long GRBs are indicated by circles, 
the short ones by squares and both time scales are with respect to the observer frame. As in the histograms above, the fact 
that short GRBs, in general, tend to have smaller $\tau_\beta$  
values compared to long GRBs, is evident in this figure. Also shown in 
the figure (as a dash line) is the trajectory of $\tau_\beta$  
equal to $T_{90}$. As we expect, no long GRBs exhibit a $\tau_\beta$  longer than 
$T_{90}$ although interestingly a few short GRBs of extremely short duration 
appear to be approaching the limit of equality. In addition to 
establishing a characteristic time scale for short and long bursts, this figure also hints at a positive correlation between this 
time and the duration of bursts. We note that the $\tau_\beta$  
scale spans approximately two decades for both sets of GRBs and that the two 
groups are fairly well clustered in the $\tau_\beta$-$T_{90}$ plane. 
A closer examination of the two groups, however, indicates that a correlation 
between $\tau_\beta$  and $T_{90}$, if present, is marginal at best. 
This is certainly true for the short-GRB group, especially given the large 
uncertainties in the $T_{90}$s for these bursts. The situation for the long-burst group 
on the other hand is not immediately clear. In 
order to explore this further we cast the $\tau_\beta$  
and the $T_{90}$ time scales into the source frame by applying the appropriate $(1+z)$ 
factor to the GRBs for which the z is known. Unfortunately the z is not available for the majority of the short GRBs but we note that 
the correction is the same for both axes and is, to first order, small for the short GRBs since the mean z for this group is $< 0.8$. 
The corrected results for long-GRBs are shown as a log-log plot in Fig.~\ref{TauVT90}.  We see from this figure (and Fig.~\ref{TauVT90Obs}) 
the appearance of a very 
intriguing feature: A plateau region in which the $\tau_\beta$  
is essentially independent of $T_{90}$ and a scaling region in which the $\tau_\beta$  
appears to increase with $T_{90}$, with the transition occurring around $T_{90}$ on the order of a few seconds.

If one assumes a positive correlation between luminosity and variability as suggested by a number of authors, then one might 
expect smaller $\tau_\beta$  values for higher luminosity bursts compared to those of lower luminosity. To investigate this, the data 
(in Fig.~\ref{TauVT90}) are re-plotted in Fig.~\ref{TauVT90Lum} 
in which the size of each datum symbol has been modulated by the gamma-ray luminosity of the burst, 
i.e., a large symbol implies a high luminosity and a small symbol a low luminosity. We see from Fig.~\ref{TauVT90Lum} 
that no obvious connection between $\tau_\beta$  and luminosity is evident.

Under the assumption that the $\tau_\beta$  is a measure proportional 
to the smallest causally-connected structure associated with a GRB 
light curve, it is then possible to interpret the scaling trend in terms of the internal shock model in which the basic units of 
emission are assumed to be pulses that are produced via the collision of relativistic shells emitted by the central engine. 
Indeed, we note that \citet{Quilligan02} in their study of the brightest BATSE bursts with $T_{90}$ $>2$ sec explicitly identified 
and fitted distinct pulses and demonstrated a strong positive correlation between the number of pulses and the duration of the burst. 
More recent studies \citep{Bhat11,Bhat12,Hakkila08,Hakkila11} provide further evidence for the pulse paradigm view of the prompt emission in GRBs. 
In our work we have not relied on identifying distinct pulses but instead have used the multi-resolution capacity of the wavelet 
technique to resolve the smallest temporal scale present in the prompt emission. If the smallest temporal scale is made from pulse 
emissions from the smallest structures, then we can get a measure of the number of pulses in a given burst through the ratio 
$T_{90}$/$\tau_\beta$ . 	
In the simple model in which a pulse is produced every time two shells collide then the ratio, 
$T_{90}$/$\tau_\beta,$ should show a correlation with 
the duration of the burst. 
A plot of this ratio versus $T_{90}$ is shown for a sample of short and long bursts in Fig.~\ref{T90BYTauVT90Obs}. 
The correlation is apparent.

It is now widely accepted that the progenitors for the two classes of GRBs are quite distinct i.e., the merger of 
compact objects in the case of short GRBs and the collapse of rapidly rotating massive stars in the case of long GRBs. 
Formation of an accretion disk in the two cases is posed in a number of models but important factors such as the size 
of the disk, the mass of the disk, the strength of the magnetic field, in addition to the magnitude of the accretion 
rate during the prompt phase, remain largely uncertain. With contributions from intrinsic variability of the 
central engine or nearby shock-wave interactions within a jet, we should not be surprised to observe a systematic 
difference in the extracted variability time scales for long and short bursts, since the progenitors have different
spatial scales. Knowing the variability timescales, we can estimate the size of an assumed emission region. From Fig.~\ref{TauVT90Obs}, 
we note that the smallest temporal-variability scale for the short bursts is approximately 3 ms and that for the long bursts is 
approximately 30 ms: These times translate to emission scales of approximately $10^8$ and $10^9$ cm respectively.
Our variability times and size scales are generally consistent with the findings of \citet{Walker00} although these authors also 
reported observing time scales as small as few microseconds. We find no evidence for variability times as low as a few microseconds.

\citet{Morsony2010} modeled the behavior of a jet propagating through the progenitor and the surrounding circumstellar material 
and showed that the resulting light curves exhibited both short-term and long–term variability. They attribute the long-term 
variability, at the scale of few seconds, to the interaction of the jet with the progenitor. The short-term scale, at the 
level of milliseconds, they attribute to the variation in the activity of the central engine itself. Alternatively, 
\citet{Zhang11} consider a model in which the prompt emission is the result of a magnetically powered outflow which 
is self-interacting and triggers rapid turbulent reconnections to power the observed GRBs. This model also predicts two 
variability components but interestingly and in contrast to the findings of \citet{Morsony2010} , it is the slow component 
that is associated with the activity of the central engine, and the fast component is linked to relativistic magnetic turbulence. 
While we are not in a position to distinguish between these two models it is intriguing nonetheless to note (see Fig.~\ref{TauVT90Obs})
that indeed there do appear to be two distinct time domains for the $\tau_\beta$: 
a plateau region dominated primarily by short bursts although 
it includes some long bursts too, and a scaling region (i.e., a hint of a correlation between $\tau_\beta$ and $T_{90}$) 
that is comprised solely of long bursts. 
In addition, we observe that the time scale in the plateau region is the order of milliseconds whereas that for the scaling 
region is approaching seconds.

There is considerable dispersion in the extracted $\tau_\beta$. The variation is evident for both short and long-duration GRBs. 
The main cause of this dispersion is not fully understood but one factor that may play a significant role is angular momentum. 
As \citet{Lindner10} note, the basic features of the prompt emission can be understood in terms of accretion that results 
via a simple ballistic infall of material from a rapidly rotating progenitor. Material with low angular momentum will radially 
accrete across the event horizon whereas the material with sufficient angular momentum will tend to circularize outside the 
innermost stable circular orbit and form an accretion disk. Simulations that go beyond the simple radial infall model 
\citep{Lindner10, Lindner11} suggest that the formation of the disk leads to an accretion shock that traverses 
outwards through the infalling material. If one assumes that the initiation of such an accretion shock and the subsequent 
emission of the prompt gamma-rays are associated with a particular time scale, the variability of this scale then 
(as given by the dispersion in $\tau_\beta$ for example) may reflect the different dynamics (initial angular momentum and 
the mass of the black hole) of each GRB in our sample. In the case of long GRBs, the mass of the central black hole can 
vary by an order of magnitude thus potentially explaining a large part of the dispersion seen in the $\tau_\beta$. However a 
similar dispersion for short bursts is difficult to reconcile using the same arguments since the mass range for the 
central black hole in standard merger models (at least for NS-NS mergers) is expected to be significantly smaller.

\section{CONCLUSIONS}

We have studied the temporal properties of a sample of prompt-emission light curves for short and long-duration GRBs detected by the 
Fermi/GBM mission. By using a technique based on wavelets we have extracted the variability timescales for these bursts. Our main 
results are summarized as follows:

a) Both short and long-duration bursts indicate a temporal variability at the level of a few milliseconds. Variability of this order 
appears to be a common feature of GRBs. This finding is consistent with the work of \citet{Walker00}. However, unlike these 
authors we do not find evidence of variability at a time scale of few microseconds.

b)  In general the short-duration bursts have a variability time scale that is significantly shorter than long-duration bursts. 
In addition, the $\tau_\beta$ values seem not to depend in any obvious way on the 
luminosity of the bursts. 
The dispersion over different 
GRBs in the extracted time scale for short-duration bursts is an order of magnitude within the smallest variability time, that 
time being approximately 3 milliseconds. 
The dispersion for the long-duration bursts is somewhat larger. The origin of the dispersion in either case is not known, 
although we should expect that the size of the initial angular momentum and the mass of the system play significant roles.
We note in passing that the 3 millisecond time scale may not be a physical lower limit and may be a result of signal to noise and the set of GRBs used in this analysis.
We remind the reader that our light curve resolution was 200 $\mu$s and if a strong enough signal within a range of time scales between 0.5 - 3 milliseconds 
were present we would expect our technique to be sensitive
to it.

c) The ratio of $T_{90}/\tau_\beta$ appears to be positively correlated with the minimum variability time scale. This suggests further
support for the pulse paradigm view of the prompt emission as being the result of shell collisions. 
In this respect, the minimum variability time scale is likely related to key pulse parameters such as risetimes and/or widths. 

d) For short-duration bursts, the variability parameter $\tau_\beta$ shows negligible dependence on the duration of the bursts 
(characterized by $T_{90}$). In contrast, the long-duration bursts indicate evidence for two variability time scales: 
a plateau region (at the shortest time scale) which is essentially independent of burst duration and a scaling region 
(at the higher time scale) that shows a positive correlation with burst duration. The transition between the two regions 
occurs around $T_{90}$ on the order of a few seconds in the source frame.

\section*{ACKNOWLEDGEMENTS}
The NASA grant NNX11AE36G provided partial support for this work and is gratefully acknowledged.
The authors, in particular GAM and KSD, acknowledge very useful discussions with Jon Hakkila and Narayan Bhat early in the manuscript development. GAM 
and KSD also acknowledge helpful correspondences with Jeffery Scargle.

%\section*{Appendix}

\begin{table}
\caption{{\small Short GRBs (Observer Frame).}}
\centering % used for centering table 
\begin{tabular}{cccccc} % centered columns (4 columns) 
\hline\hline %inserts double horizontal lines 
GRB & $T_{90}$[sec] & $\delta T_{90}$[sec] & $\tau_{\beta}$ [sec] & $\delta\tau^-_{\beta}$ [sec] & $\delta\tau^+_{\beta}$ [sec] \\
\hline
080723913 & 0.192 & 0.345 & 0.0307 & 0.0192 & 0.0510 \\ 
081012045 & 1.216 & 1.748 & 0.0052 & 0.0024 & 0.0044 \\ 
081102365 & 1.728 & 0.231 & 0.0258 & 0.0100 & 0.0165 \\ 
081105614 & 1.280 & 1.368 & 0.0306 & 0.0147 & 0.0282 \\ 
081107321 & 1.664 & 0.234 & 0.0504 & 0.0129 & 0.0173 \\ 
081216531 & 0.768 & 0.429 & 0.0138 & 0.0037 & 0.0050 \\ 
090108020 & 0.704 & 0.143 & 0.0241 & 0.0064 & 0.0088 \\ 
090206620 & 0.320 & 0.143 & 0.0143 & 0.0063 & 0.0112 \\ 
090227772 & 1.280 & 1.026 & 0.0053 & 0.0009 & 0.0011 \\ 
090228204 & 0.448 & 0.143 & 0.0028 & 0.0005 & 0.0005 \\ 
090308734 & 1.664 & 0.286 & 0.0120 & 0.0040 & 0.0059 \\ 
090429753 & 0.640 & 0.466 & 0.0285 & 0.0115 & 0.0193 \\ 
090510016 & 0.960 & 0.138 & 0.0049 & 0.0009 & 0.0011 \\ 
100117879 & 0.256 & 0.834 & 0.0331 & 0.0122 & 0.0192 \\ 
\hline %inserts single line
\hline 
\end{tabular}
\label{table:shorts} % is used to refer this table in the text 
\end{table}
\onecolumn
\begin{table}
\caption{{\small Long GRBs (Observer Frame).}}
\centering % used for centering table 
\small{
\begin{tabular}{cccccc} % centered columns (4 columns) 
\hline\hline %inserts double horizontal lines 
GRB & $T_{90}$[sec] & $\delta T_{90}$[sec] & $\tau_{\beta}$ [sec] & $\delta\tau^-_{\beta}$ [sec] & $\delta\tau^+_{\beta}$ [sec] \\
\hline
080723557 & 58.369 & 1.985 & 0.0440 & 0.0113 & 0.0151 \\ 
080723985 & 42.817 & 0.659 & 0.1894 & 0.0557 & 0.0789 \\ 
080724401 & 379.397 & 2.202 & 0.0741 & 0.0208 & 0.0290 \\ 
080804972 & 24.704 & 1.460 & 0.4306 & 0.1336 & 0.1937 \\ 
080806896 & 75.777 & 4.185 & 0.4189 & 0.1471 & 0.2268 \\ 
080807993 & 19.072 & 0.181 & 0.0232 & 0.0096 & 0.0164 \\ 
080810549 & 107.457 & 15.413 & 0.1353 & 0.0648 & 0.1243 \\ 
080816503 & 64.769 & 1.810 & 0.1067 & 0.0428 & 0.0715 \\ 
080817161 & 60.289 & 0.466 & 0.1919 & 0.0402 & 0.0509 \\ 
080825593 & 20.992 & 0.231 & 0.0775 & 0.0138 & 0.0168 \\ 
080906212 & 2.875 & 0.767 & 0.1011 & 0.0182 & 0.0222 \\ 
080916009 & 62.977 & 0.810 & 0.2266 & 0.0630 & 0.0872 \\ 
080925775 & 31.744 & 3.167 & 0.1748 & 0.0425 & 0.0562 \\ 
081009140 & 41.345 & 0.264 & 0.1095 & 0.0170 & 0.0201 \\ 
081101532 & 8.256 & 0.889 & 0.0948 & 0.0302 & 0.0444 \\ 
081125496 & 9.280 & 0.607 & 0.2182 & 0.0504 & 0.0656 \\ 
081129161 & 62.657 & 7.318 & 0.0912 & 0.0292 & 0.0429 \\ 
081215784 & 5.568 & 0.143 & 0.0319 & 0.0043 & 0.0050 \\ 
081221681 & 29.697 & 0.410 & 0.2701 & 0.0641 & 0.0841 \\ 
081222204 & 18.880 & 2.318 & 0.1956 & 0.0533 & 0.0732 \\ 
081224887 & 16.448 & 1.159 & 0.2055 & 0.0356 & 0.0431 \\ 
090102122 & 26.624 & 0.810 & 0.0347 & 0.0111 & 0.0164 \\ 
090131090 & 35.073 & 1.056 & 0.0733 & 0.0169 & 0.0220 \\ 
090202347 & 12.608 & 0.345 & 0.1444 & 0.0575 & 0.0954 \\ 
090323002 & 135.170 & 1.448 & 0.1598 & 0.0436 & 0.0599 \\ 
090328401 & 61.697 & 1.810 & 0.0682 & 0.0139 & 0.0175 \\ 
090411991 & 14.336 & 1.086 & 0.0673 & 0.0391 & 0.0935 \\ 
090424592 & 14.144 & 0.264 & 0.0249 & 0.0031 & 0.0036 \\ 
090425377 & 75.393 & 2.450 & 0.1346 & 0.0369 & 0.0508 \\ 
090516137 & 118.018 & 4.028 & 0.4938 & 0.2063 & 0.3544 \\ 
090516353 & 123.074 & 2.896 & 0.7992 & 0.5686 & 1.9711 \\ 
090528516 & 79.041 & 1.088 & 0.1314 & 0.0320 & 0.0423 \\ 
090618353 & 112.386 & 1.086 & 0.2631 & 0.0536 & 0.0673 \\ 
090620400 & 13.568 & 0.724 & 0.1667 & 0.0422 & 0.0564 \\ 
090626189 & 48.897 & 2.828 & 0.0498 & 0.0078 & 0.0093 \\ 
090718762 & 23.744 & 0.802 & 0.1621 & 0.0482 & 0.0686 \\ 
090809978 & 11.008 & 0.320 & 0.2436 & 0.0515 & 0.0652 \\ 
090810659 & 123.458 & 1.747 & 0.7319 & 0.3027 & 0.5161 \\ 
090829672 & 67.585 & 2.896 & 0.0678 & 0.0141 & 0.0177 \\ 
090831317 & 39.424 & 0.572 & 0.0266 & 0.0103 & 0.0169 \\ 
090902462 & 19.328 & 0.286 & 0.0223 & 0.0026 & 0.0029 \\ 
090926181 & 13.760 & 0.286 & 0.0435 & 0.0061 & 0.0070 \\ 
091003191 & 20.224 & 0.362 & 0.0300 & 0.0051 & 0.0062 \\ 
091127976 & 8.701 & 0.571 & 0.0395 & 0.0059 & 0.0069 \\ 
091208410 & 12.480 & 5.018 & 0.0621 & 0.0180 & 0.0254 \\ 
100414097 & 26.497 & 2.073 & 0.0418 & 0.0074 & 0.0090 \\ 
\hline %inserts single line 
\hline
\end{tabular}
}
\label{table:longs} % is used to refer this table in the text 
\end{table}
\twocolumn
\onecolumn
\begin{table}
\caption{{\small Long and Short GRBs ($T_{90}$ and $\tau_\beta$ in Observer Frame).  Luminosities are taken from references given in footnotes.  }}
\centering % used for centering table 
\begin{tabular}{cccccccccc} % centered columns (4 columns) 
\hline\hline %inserts double horizontal lines 
GRB  & $z$ & $T_{90}$[sec] & $\delta T_{90}$[sec] & $\tau_{\beta}$ [sec] & $\delta\tau^-_{\beta}$ [sec] & $\delta\tau^+_{\beta}$ [sec]
&  $L_{iso}$ [ergs/s]  & $\delta L_{iso}^-$ [ergs/s] & $\log$ $\delta L_{iso}^+$ [ergs/s]   \\
\hline
080804972 & 2.204 & 24.704 & 1.460 & 0.4306 & 0.1336 & 0.1937 & \footnotemark[1]3.58$\cdot 10^{52}$ & 5.82$\cdot 10^{51}$ & 7.85$\cdot 10^{51}$ \\ 
080810549 & 3.350 & 107.457 & 15.413 & 0.1353 & 0.0648 & 0.1243 & \footnotemark[2]9.59$\cdot 10^{52}$ & 1.28$\cdot 10^{52}$ & 1.28$\cdot 10^{52}$ \\ 
080916009 & 4.350 & 62.977 & 0.810 & 0.2266 & 0.0630 & 0.0872 & \footnotemark[3]1.04$\cdot 10^{54}$ & 8.79$\cdot 10^{52}$ & 8.79$\cdot 10^{52}$ \\ 
081222204 & 2.770 & 18.880 & 2.318 & 0.1956 & 0.0533 & 0.0732 & \footnotemark[1]1.26$\cdot 10^{53}$ & 7$\cdot 10^{51}$ & 6$\cdot 10^{51}$ \\ 
090102122 & 1.547 & 26.624 & 0.810 & 0.0347 & 0.0111 & 0.0164 & \footnotemark[3]8.71$\cdot 10^{52}$  & 5.6$\cdot 10^{51}$ & 5.6$\cdot 10^{51}$ \\ 
090323002 & 3.570 & 135.170 & 1.448 & 0.1598 & 0.0436 & 0.0599 & \footnotemark[1]6.87$\cdot 10^{53}$  & 6.55$\cdot 10^{53}$ & 4.45$\cdot 10^{52}$ \\ 
090328401 & 0.736 & 61.697 & 1.810 & 0.0682 & 0.0139 & 0.0175 & \footnotemark[4]1.79$\cdot 10^{52}$  & 1.42$\cdot 10^{51}$ & 1.11$\cdot 10^{51}$ \\ 
090424592 & 0.544 & 14.144 & 0.264 & 0.0249 & 0.0031 & 0.0036 & \footnotemark[5]1.62$\cdot 10^{52}$  & 4$\cdot 10^{50}$ & 5$\cdot 10^{50}$ \\ 
090510016 & 0.903 & 0.960 & 0.138 & 0.0049 & 0.0009 & 0.0011 & \footnotemark[3]1.78$\cdot 10^{53}$  & 1.2$\cdot 10^{51}$ & 1.2$\cdot 10^{51}$ \\ 
090516353 & 4.100 & 123.074 & 2.896 & 0.7992 & 0.5686 & 1.9711 & \footnotemark[6]8.17$\cdot 10^{52}$  & 2.85$\cdot 10^{52}$ & 6.1$\cdot 10^{51}$ \\ 
090618353 & 0.540 & 112.386 & 1.086 & 0.2631 & 0.0536 & 0.0673 & \footnotemark[5]8.47$\cdot 10^{51}$  & 1.17$\cdot 10^{51}$ & 3.4$\cdot 10^{50}$ \\ 
090902462 & 1.822 & 19.328 & 0.286 & 0.0223 & 0.0026 & 0.0029 & \footnotemark[3]5.89$\cdot 10^{53}$  & 9.71$\cdot 10^{51}$ & 9.71$\cdot 10^{51}$ \\ 
090926181 & 2.106 & 13.760 & 0.286 & 0.0435 & 0.0061 & 0.0070 & \footnotemark[3]7.40$\cdot 10^{53}$ & 1.45$\cdot 10^{52}$ & 1.45$\cdot 10^{52}$ \\ 
091003191 & 0.897 & 20.224 & 0.362 & 0.0300 & 0.0051 & 0.0062 & \footnotemark[1]4.53$\cdot 10^{52}$  & 3.71$\cdot 10^{51}$ & 6.55$\cdot 10^{51}$ \\ 
091127976 & 0.490 & 8.701 & 0.571 & 0.0395 & 0.0059 & 0.0069 & \footnotemark[7]3.70$\cdot 10^{51}$  & 1.38$\cdot 10^{50}$ & 1.06$\cdot 10^{50}$ \\ 
091208410 & 1.063 & 12.480 & 5.018 & 0.0621 & 0.0180 & 0.0254 & \footnotemark[1]1.45$\cdot 10^{52}$  & 1.48$\cdot 10^{51}$ & 3.45$\cdot 10^{51}$ \\ 
100117879 & 0.920 & 0.256 & 0.834 & 0.0331 & 0.0122 & 0.0192 & \footnotemark[1]2.63$\cdot 10^{52}$  & 5.01$\cdot 10^{51}$ & 1.08$\cdot 10^{52}$ \\ 
100414097 & 1.368 & 26.497 & 2.073 & 0.0418 & 0.0074 & 0.0090 & \footnotemark[8]1.00$\cdot 10^{53}$  & 1.58$\cdot 10^{52}$ & 7.6$\cdot 10^{51}$ \\ 
\hline
\hline %inserts single line 
\end{tabular}
\label{table:good} % is used to refer this table in the text 
\end{table}
\footnotetext[1]{Nava et al. 2011 (A\&A 530, A21 (2011))}
\footnotetext[2]{GCN \#8100}
\footnotetext[3]{Ghirlanda et al. 2011  (arXiv:1107.4096)}
\footnotetext[4]{GCN \#9057}
\footnotetext[5]{Ukwatta, T.~N., et al. 2010, \apj, 711, 1073} 
\footnotetext[6]{GCN \#9415}
\footnotetext[7]{GCN \#10204}
\footnotetext[8]{GCN \#10595}

\twocolumn

\end{document}